\documentclass[
    ,final            % use final for the camera ready runs
  ]
  {aipproc}

\layoutstyle{6x9}

\begin{document}

\title{The Galactic Center Region Imaged by VERITAS from 2010--2012}

%-----------------[ Dummy
%http://www.aip.org/pacs/index.html


\author{M.~Beilicke}{address={Department of Physics and MCSS, Washington
University, St. Louis, MO, USA}, email={beilicke@physics.wustl.edu}}

\author{the VERITAS
Collaboration}{address={\url{http://veritas.sao.arizona.edu/}}}

\begin{abstract}

The galactic center (GC) has long been a region of interest for
high-energy and very-high-energy observations. Many potential sources of
GeV/TeV $\gamma$-ray emission are located in the GC region, e.g. the
accretion of matter onto the central black hole (BH), cosmic rays from a
nearby shell-type super nova remnant, or the annihilation of dark
matter. The GC has been detected at MeV/GeV energies by EGRET and
recently by {\it Fermi}/LAT. At TeV energies, the GC was detected at the
level of 4 standard deviations with the Whipple $10 \, \rm{m}$ telescope
and with one order of magnitude better sensitivity by H.E.S.S. and
MAGIC.  We present the results from 3 years of VERITAS GC observations
conducted at large zenith angles (LZA). The results are compared to
astrophysical models.

\end{abstract}

\maketitle

%%%%%%%%%%%%%%%%%%%%%%%%%%%%%%%%%%%%%%%%%%%%
%%%%%%%%%%%%%%%%%%%%%%%%%%%%%%%%%%%%%%%%%%%%
%% MAINMATTER
%%%%%%%%%%%%%%%%%%%%%%%%%%%%%%%%%%%%%%%%%%%%
%%%%%%%%%%%%%%%%%%%%%%%%%%%%%%%%%%%%%%%%%%%%

%% ############################################################
%% ############ Introduction
%% ############################################################
\section{Introduction}
\label{sec:Introduction}

The center of our galaxy harbors a $4 \times 10^{6} \, M_{\odot}$ BH
coinciding with the strong radio source Sgr\,A*. X-ray/MeV/GeV
transients in this region are observed on a regular basis. Various
astrophysical sources are located close to the GC which may potentially
be capable of accelerating particles to multi-TeV energies, such as the
supernova remnant Sgr\,A~East or a pulsar wind nebula \cite{GC_Plerion}.
Furthermore, the super-symmetric neutralinos $\chi$ are discussed as
potential candidates of dark matter accumulating in the GC region and
annihilating into $\gamma$-rays \cite{Neutralino}. The resulting
spectrum would have a cut-off near the neutralino mass $m_{\chi}$.
Assuming a certain dark matter density profile the expected $\gamma$-ray
flux along the line-of-sight integral can be calculated as a function of
$m_{\chi}$ and the annihilation cross section \cite{GammasFromNWF} and
can in turn be compared to measured upper limits.

EGRET detected a MeV/GeV source 3EG\,J1746-2851 coincident with the GC
position \cite{Egret_GC} and recently {\it Fermi}/LAT resolved several
sources in the GC region \cite{Fermi_FirstCatalog}, see
Fig.~\ref{fig:Skymap}. However, uncertainties in the diffuse galactic
background models and limited angular resolution at MeV/GeV make it
difficult to study the morphologies of these sources. At GeV/TeV
energies a detection from the direction of the GC was first reported in
2001/02 by the CANGAROO\,II collaboration with a steep energy spectrum
$\rm{d}N/\rm{d}E \propto E^{-4.6}$ at the level of $10\%$ of the Crab
Nebula flux \cite{CANGAROO_GC}. Shortly after, evidence at the level of
$3.7$ standard deviations (s.d.) was reported from the Whipple $10 \,
\rm{m}$ collaboration \cite{Whipple_GC}. The GC was finally confirmed as
a GeV/TeV $\gamma$-ray source by the H.E.S.S. collaboration
\cite{HESS_SgrA} (the position of the supernova remnant Sgr\,A~East
could be excluded as the source of the $\gamma$-ray emission). The
energy spectrum is well described by a power-law $\rm{d}N/\rm{d}E
\propto E^{-2.1}$ with a cut-off at $\sim$$15 \, \rm{TeV}$. The
H.E.S.S.~observations revealed a diffuse GeV/TeV $\gamma$-ray component
(dashed contour lines in Fig.~\ref{fig:Skymap}) which is aligned along
the galactic plane and follows the structure of molecular clouds
\cite{HESS_SgrA_Diffuse}; the emission was explained by an interaction
of local cosmic rays (CRs) with matter of the molecular clouds. The
MAGIC collaboration detected the GC (7~s.~d.) in 2004/05 observations
performed at LZA \cite{MAGIC_GC}, followed by a strong VERITAS LZA
detection in 2010 \cite{Beilicke2011}.

%% ############################################################
%% ############ VERITAS observations of the galactic center
%% ############################################################
\section{The Galactic Center region imaged by VERITAS}

%-------------
\begin{figure}[t]

\includegraphics[width=0.99\textwidth]{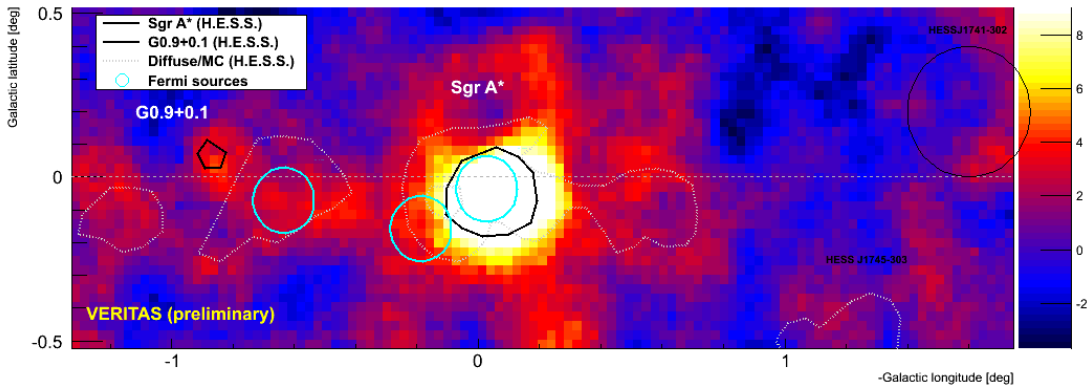}

\caption{\label{fig:Skymap} VERITAS sky map of the GC region (smoothed
excess significances, ring background, scale truncated). The black
contour lines indicate the GC and the supernova remnant G\,0.9+0.1 as
seen by H.E.S.S. \cite{HESS_SgrA}. The gray dashed lines indicate the
H.E.S.S. diffuse emission along the galactic plane and from
HESS\,J1745-303 \cite{HESS_SgrA_Diffuse}. The position of
HESS\,J1741-302 is indicated, as well (circle). The solid circles (cyan
color) indicate the positions of the MeV/GeV sources taken from the
second {\it Fermi}/LAT catalog \cite{Fermi_FirstCatalog}.}

\end{figure}

\paragraph{GC observations} Due to its declination the GC can only be
observed by VERITAS at LZA ($z = 60-66 \deg$)~-- strongly decreasing the
angular resolution and sensitivity. The use of the {\it displacement}
parameter \cite{BuckelyDisp}, between the center of gravity of the image
and the shower position, has been used in the VERITAS event
reconstruction which strongly improved the sensitivity for LZA
observations \cite{Beilicke2011}. The performance and energy
reconstruction have been confirmed on LZA Crab Nebula data. The column
density of the atmosphere changes with $1 / \cos(z)$. In a conservative
estimate, the systematic error in the energy/flux reconstruction can be
expected to scale accordingly.  More detailed studied are needed for an
accurate estimate; for the GC observations we currently give a
conservative value of a systematic error on the LZA flux normalization
of $\Delta \Phi / \Phi \simeq 0.4$. The GC was observed by VERITAS in
2010--2012 for $46 \, \rm{hrs}$ (good quality data, dead-time corrected)
with an average energy threshold of $E_{\rm{thr}} \simeq 2.5 \,
\rm{TeV}$. 

\begin{figure*}[t]

\includegraphics[width=0.49\textwidth]{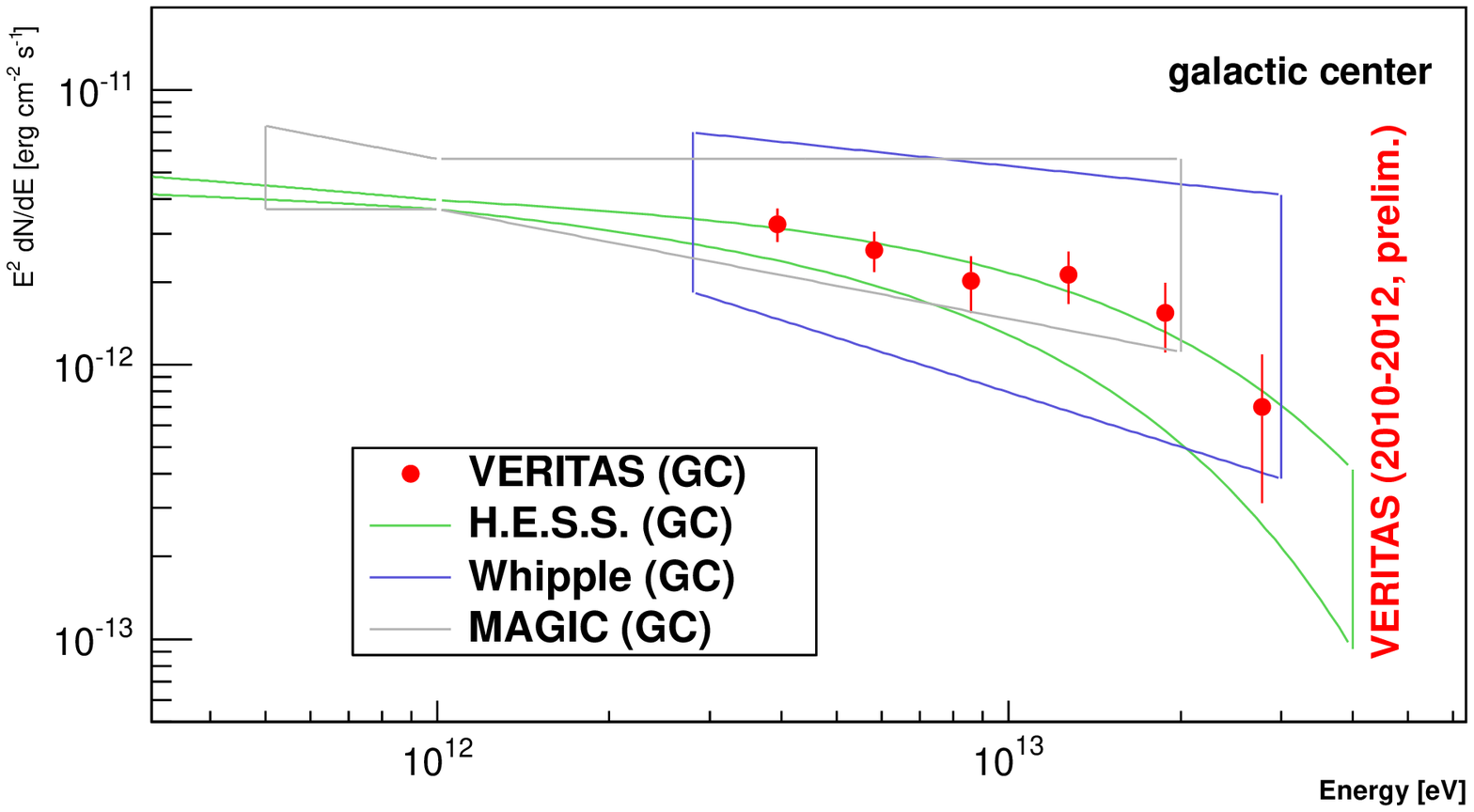}
\hfill
\includegraphics[width=0.49\textwidth]{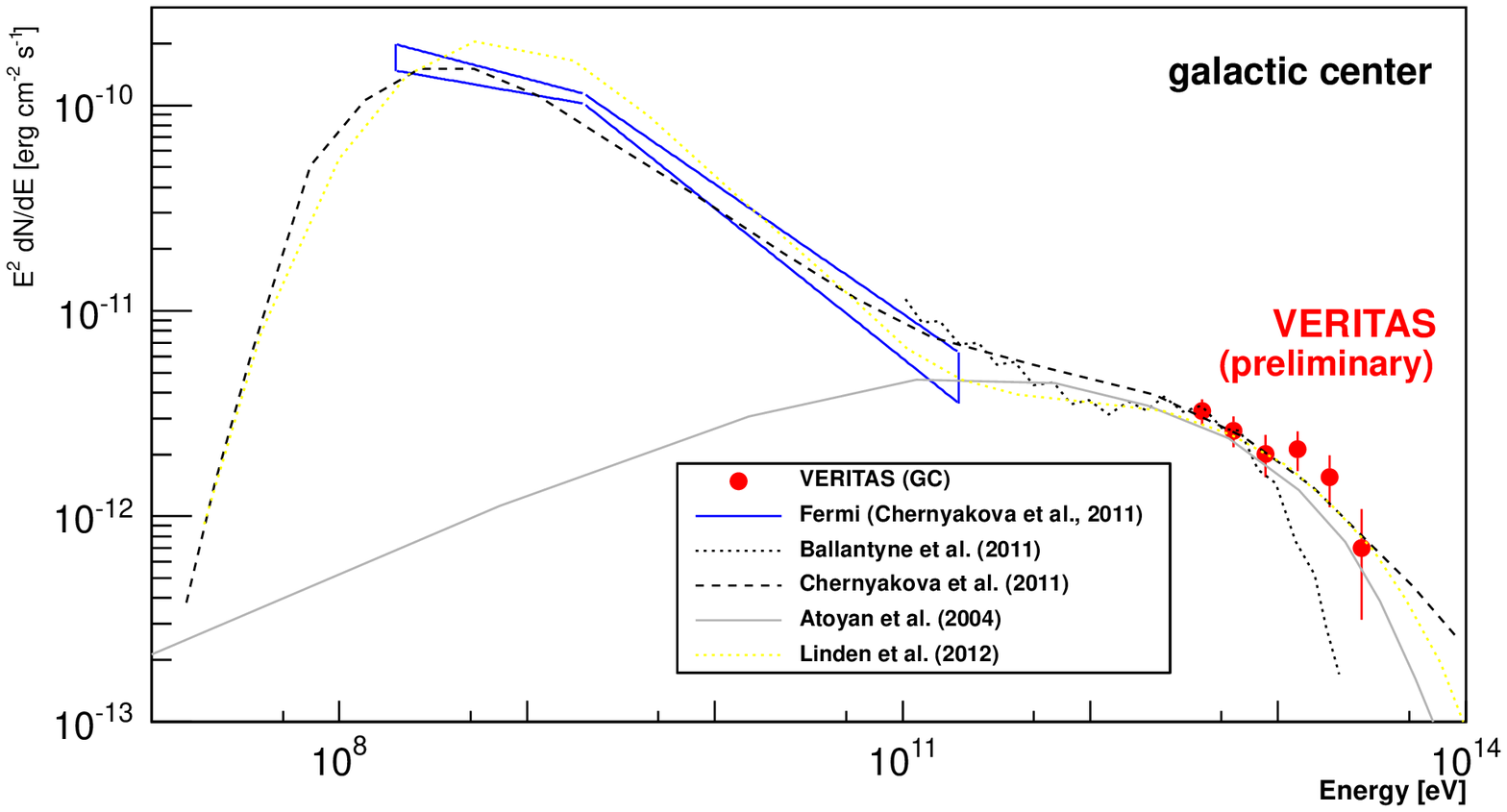}

\caption{\label{fig:SED} {\bf Left:} VERITAS energy spectrum measured
from the direction of the GC (statistical errors only). Also shown are
bow ties representing the spectra measured by Whipple \cite{Whipple_GC},
H.E.S.S. \cite{HESS_SgrA}, and MAGIC \cite{MAGIC_GC}. {\bf Right:}
VERITAS energy spectrum compared to hadronic \cite{Chernyakova2011,
Ballantyne2011, Linden2012} and leptonic \cite{Atoyan2004} emission
models discussed for the GC source. The {\it Fermi}/LAT bow tie is taken
from \cite{Chernyakova2011}.}

\end{figure*}

\paragraph{GC results} The VERITAS sky map of the GC region is shown in
Fig.~\ref{fig:Skymap}. An $18$~s.d.~excess is detected. No evidence for
variability was found in the 3-year data. The energy spectrum is shown
in Fig.~\ref{fig:SED} and is found to be compatible with the spectra
measured by Whipple, H.E.S.S., and MAGIC. Since the large LZA effective
areas of the VERITAS observations compensate a shorter exposure of
low-zenith observations, the statistical errors of the $E > 2.5 \,
\rm{TeV}$ data points are found to be smaller as compared to the
H.E.S.S.~measurements.

\paragraph{Diffuse flux limit and dark-matter annihilation} OFF-source
observations were performed in a field located in the vicinity of the GC
region (similar zenith angles and sky brightness) without a known TeV
$\gamma$-ray source. These observations are used to study the background
acceptance throughout the field of view and will allow the estimate of a
diffuse $\gamma$-ray component surrounding the position of the GC. An
upper limit of the diffuse $\gamma$-ray flux can in turn be compared
with line-of-sight integrals along the density profile $\int \rho^{2}
\rm{d}l$, in order to constrain the annihilation cross section for a
particular dark matter model, dark matter particle mass and density
profile $\rho(r)$. Due to its likely astrophysical origin the excess at
the GC itself, as well as a region along the galactic plane, will be
excluded from this analysis (work in progress).

\paragraph{Hadronic models} Hadronic acceleration models
\cite{Chernyakova2011, Ballantyne2011} are discussed involving: (i)
hadrons being accelerated in the BH vicinity (few tens of Schwarzschild
radii). (ii) The accelerated protons diffuse out into the interstellar
medium where they (iii) produce neutral pions which decay into GeV/TeV
$\gamma$-rays.  Linden et al. (2012) discuss the surrounding gas as
proton target defining the morphology of the TeV $\gamma$-ray emission
\cite{Linden2012}. Changes in $\gamma$-ray flux in those models can be
caused by changing conditions in the BH vicinity (e.g. accretion). The
time scales of flux variations are $\sim$$10^{4} \, \rm{yr}$ at MeV/GeV
energies (old flares) and $\sim$$10 \, \rm{yr}$ at $E > 10 \, \rm{TeV}$
('new' flares caused by recently injected high-energy particles)
\cite{Chernyakova2011}. Constraining the $E > 10 \, \rm{TeV}$ spectral
variability would serve as an important test for this class of models.

%-------------

\paragraph{Leptonic models} Atoyan et al. (2004) \cite{Atoyan2004}
discuss a BH plerion model in which a termination shock of a leptonic
wind accelerates leptons to relativistic energies which in turn produce
TeV $\gamma$-rays via inverse Compton scattering. The flux variability
time scale in this model is on the order of $T_{\rm{var}} \sim$$100 \,
\rm{yr}$. The hadronic and the leptonic models are shown together with
the VERITAS/{\it Fermi} data in Fig.~\ref{fig:SED} (right). The leptonic
model clearly fails in explaining the flux in the MeV/GeV regime.
However, this emission may well originate from a spatially different
region or mechanism other than the TeV $\gamma$-ray emission. The
hadronic models can explain the SED by the superposition of different
flare stages. Future {\it Fermi}/VERITAS flux correlation studies, as
well as the measurement of the TeV energy cut-off and limits on the $E >
10 \, \rm{TeV}$ variability will serve as crucial inputs for the
modeling.

%% ############################################################
%% ############ Summary and Conclusion
%% ############################################################
\section{Summary and conclusion}

VERITAS is capable of detecting the GC within $3 \, \rm{hrs}$ in zenith
angle greater than $60 \, \deg$ observations. The measured energy
spectrum is found to be in agreement with earlier measurements by
H.E.S.S., MAGIC, and Whipple.  Future observations to measure the
cut-off energy in the spectrum and to determine limits on the flux
variability at the highest energies will place constraints on emission
models. The recently discovered giant molecular cloud heading towards
the immediate vicinity of the GC BH \cite{BH_eats_MC} represents further
motivation for future TeV $\gamma$-ray monitoring of this region.  An
upper limit on diffuse $\gamma$-ray emission and, in consequence, a
limit on the photon flux initiated by the annihilation of dark matter
particles is work in progress.

%%%%%%%%%%%%%%%%%%%%%%%%%%%%%%%%%%%%%%%%%%%%%%%%
%% Acknowledgments
%%%%%%%%%%%%%%%%%%%%%%%%%%%%%%%%%%%%%%%%%%%%%%%%

\begin{theacknowledgments}

This research is supported by grants from the U.S. Department of Energy
Office of Science, the U.S. National Science Foundation and the
Smithsonian Institution, by NSERC in Canada, by Science Foundation
Ireland (SFI 10/RFP/AST2748) and by STFC in the U.K. We acknowledge the
excellent work of the technical support staff at the Fred Lawrence
Whipple Observatory and at the collaborating institutions in the
construction and operation of the instrument.

\end{theacknowledgments}


\bibliographystyle{aipproc}   % if natbib is available

%\bibliography{sample}

%%%%%%%%%%%%%%%%%%%%%%%%%%%%%%%%%%%%%%%%%%%
%% The Bibliography
%%%%%%%%%%%%%%%%%%%%%%%%%%%%%%%%%%%%%%%%%%%

\end{document}